\documentclass[aps,prb,twocolumn,showpacs,groupedaddress]{revtex4-1}
\usepackage{graphicx}

\newcommand*{\citen}[1]{%
  \begingroup
    \romannumeral-`\x 
    \setcitestyle{numbers}%
    \cite{#1}%
  \endgroup   
}

\begin{document}

\title{Origin of the turn-on temperature behavior in WTe$_2$}

\author{Y. L. Wang$^{1,2}$}
\author{L. R. Thoutam$^{1,3}$}
\author{Z. L. Xiao$^{1,3}$}\email{xiao@anl.gov or zxiao@niu.edu}
\author{J. Hu$^{4}$}
\author{S. Das$^{5}$}
\author{Z. Q. Mao$^{4}$}
\author{J. Wei$^{4}$}
\author{R. Divan$^{5}$}
\author{A. Luican-Mayer$^{5}$}
\author{G. W. Crabtree$^{1,6}$}
\author{W. K. Kwok$^{1}$}

\affiliation{$^{1}$Materials Science Division, Argonne National Laboratory, Argonne, Illinois 60439, USA}

\affiliation{$^{2}$Department of Physics, University of Notre Dame, Notre Dame, Indiana 46556, USA}

\affiliation{$^{3}$Department of Physics, Northern Illinois University, DeKalb, Illinois 60115, USA}

\affiliation{$^{4}$Department of Physics and Engineering Physics, Tulane University, New Orleans, Louisiana 70118, USA}

\affiliation{$^{5}$Center for Nanoscale Materials, Argonne National Laboratory, Argonne, Illinois 60439, USA}

\affiliation{$^{6}$Departments of Physics, Electrical and Mechanical Engineering, University of Illinois at Chicago, Illinois 60607, USA}

\date{\today}

\begin{abstract}
A hallmark of materials with extremely large magnetoresistance (XMR) is the transformative ‘turn-on’ temperature behavior: when the applied magnetic field $H$ is above certain value, the resistivity versus temperature $\rho(T)$ curve shows a minimum at a field dependent temperature $T^*$, which has been interpreted as a magnetic-field-driven metal-insulator transition or attributed to an electronic structure change. Here, we demonstrate that $\rho(T)$ curves with turn-on behavior in the newly discovered XMR material WTe$_2$ can be scaled as MR $\sim(H/\rho_0)^m$ with $m\approx 2$ and $\rho_0$ being the resistivity at zero-field. We obtained experimentally and also derived from the observed scaling the magnetic field dependence of the turn-on temperature $T^* \sim (H-H_c)^\nu$ with $\nu \approx 1/2$, which was earlier used as evidence for a predicted metal-insulator transition. The scaling also leads to a simple quantitative expression for the resistivity $\rho^* \approx 2 \rho_0$ at the onset of the XMR behavior, which fits the data remarkably well. These results exclude the possible existence of a magnetic-field-driven metal-insulator transition or significant contribution of an electronic structure change to the low-temperature XMR in WTe$_2$. This work resolves the origin of the turn-on behavior observed in several XMR materials and also provides a general route for a quantitative understanding of the temperature dependence of MR in both XMR and non-XMR materials.
\end{abstract}


\maketitle

The electrical resistance of a material can change its value in the presence of a magnetic field.\cite{R1,R2,R3,R4,R5,R6,R7,R8,R9,R10,R11,R12,R13,R14,R15,R16,R17,R18,R19,R20,R21,R22,R23,R24,R25,R26,R27,R28,R29,R30,R31} Such magnetic-field-induced resistance changes - the magnetoresistance (MR) – are at the core of hard drives in computers\cite{R7} and other applications such as magnetic field sensors.\cite{R32,R33} Since larger MRs give rise to higher sensitivities of the devices, searching for new materials with large MRs has perpetually remained at the frontier of contemporary materials science research.\cite{R5,R6,R9,R13,R14,R15,R16,R17,R18,R19,R20,R21,R22,R23,R24,R25,R26,R27,R28,R29,R30,R31} Besides the giant MR (GMR)\cite{R7} and colossal MR (CMR)\cite{R4} found in magnetic thin films and compounds, extremely large MR (XMR) has been revealed in graphite,\cite{R9,R10} bismuth\cite{R6} and many non-magnetic compounds such as PtSn\cite{R4,R13} PdCoO$_2$,\cite{R14} WTe$_2$,\cite{R15,R16,R17,R18,R19,R20,R21} NbSb$_2$,\cite{R22}  as well as the newly discovered 3D Dirac semimetals Cd$_3$As$_2$,\cite{R23} Na$_3$Bi,\cite{R24,R25} and topological Weyl semimetals NbP,\cite{R26,R27} NbAs,\cite{R28,R29} and TaAs.\cite{R30,R31} Among them, the recently discovered XMR in WTe$_2$ can reach $13$ million percent at $0.53$ K and in a field of $60$ T.\cite{R15,R16} 

A unique feature in the magnetoresistance of all XMR materials is its transformative 'turn-on' temperature behavior:\cite{R15,R16} when the applied magnetic field is above certain value, the resistivity
versus temperature $\rho(T)$ curve shows a minimum at a field dependent temperature $T^*$. At $T < T^*$, the resistivity increases dramatically with decreasing temperature while at $T > T^*$, it has a similar metallic temperature dependence as that at zero-field. Such a marked up-turn behavior has been
commonly attributed to a magnetic-field-driven metal-insulator transition,\cite{R17,R19,R20,R21,R22,R34,R35} although few publications questioned such an interpreattion.\cite{R10,R36,R37} In WTe$_2$, an electronic structure change has also been proposed to be a possible origin.\cite{R38,R39,R40,R41,R42} Here we tackle the issue of the MR’s turn-on temperature behavior. We demonstrate that Kohler's rule scaling, $\mathrm{MR} \sim (H/\rho_0)^m$ with $m \approx 2$, could explain the remarkable up-turn behavior in WTe$_2$. Furthermore, we discuss its universality and applicability to XMR materials with closely balanced hole and electron densities as well as other systems where one type of charge carriers may be dominant.

We measured two samples extracted from crystals grown using a chemical vapor transport method similar to that described in Refs.\citen{R15,R16} and  \citen{R43}. Four-probe dc resistive measurements were carried out in a Quantum Design PPMS-9 using a constant current mode. The magnetic field is applied along the c-axis of the crystal and is perpendicular to the current $I$ which flows along the a-axis of the crystal (see Fig.S1 for contact configuration and Table S1 for more
sample parameters). In order to more accurately determine sample temperature, the resistivity versus temperature $\rho (T)$ curves at various magnetic fields were constructed by measuring $\rho (H)$ at various fixed temperatures. More measurement details can be found in Ref.\cite{R39}.

Following the conventional definition,\cite{R13,R14,R15} we present the magnetoresistance as $\mathrm{MR} = [\rho_{xx}(T,H)-\rho_{xx}(T,0)]/\rho_{xx}(T,0)$ ($\rho_{xx}$ is the longitudinal resistivity; $\rho_{xx}(T,0)$ is also presented as $\rho_0$, the standard convention used in the literature). Figure 1 shows the typical temperature behavior of our WTe$_2$ crystals: in the absence of an external magnetic field, the resistivity deceases monotonically with
temperature. At low magnetic fields ($\leq 0.5$ T) the temperature behavior remains metallic throughout the entire temperature range. When the magnetic field is ramped to $1$ T and above, a turn-on behavior occurs, whereby at low temperatures the resistivity increases with decreasing temperature, resulting in a resistivity minimum at $T^*$. Intuitively, such a temperature behavior
can be a direct consequence of a metal-insulator transition. In analogy with the phenomenon of dynamical chiral symmetry breaking in the relativistic theories of the $(2+1)$-dimensional Dirac fermions, Khveshchenko\cite{R34} predicted that an external magnetic field can open an excitonic gap in the linear spectrum of the Coulomb interacting quasiparticles in graphite. Consequently, the
temperature at which the gap becomes zero follows a relationship of $T^* \sim H^{1/2}$, which can account for the experimental finding of $T^* \sim (H-H_c)^\nu$ with $\nu \approx 1/2$ in both graphite and bismuth, apart from the offset field $H_c$.\cite{R34,R35} Although there is no quantitative analysis on the $T^*(H)$,\cite{R15,R16,R19} the turn-on behavior in WTe$_2$ has also been interpreted as a metal-insulator transition.\cite{R17,R19,R42} As
shown in Fig.S2a, for our WTe$_2$ crystal in various fixed magnetic fields, however, the MRs increase monotonically with decreasing temperature. Although the MR does increase faster at temperatures below $T^*$, no radical changing feature such as a step from a possible gap-opening can be identified at $T^*$ in the MR($T$) curves. As plotted in Fig.S2b, the MR curves obtained at different magnetic fields can in fact overlap each other if they are normalized with the values at $5$
K. That is, the MRs at different magnetic fields have the same temperature dependence, differing from the expected behavior induced by excitonic gaps that should result in a faster change rate at a higher magnetic field. This implies that a metal-insulator transition is probably not the origin of the turn-on temperature behavior in our WTe$_2$ crystals, consistent with the observation reported by us in commercial WTe$_2$ crystals.\cite{R39}

 \begin{figure}
 \includegraphics[width=0.42\textwidth]{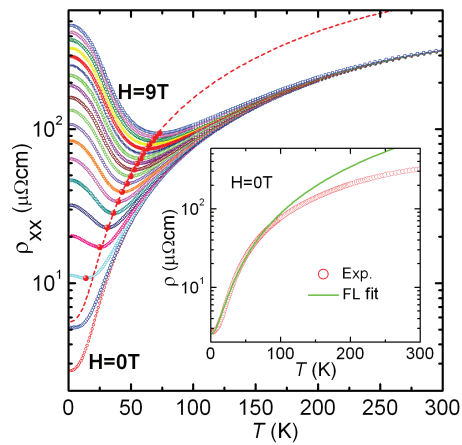}%
 \caption{\label{fig:fig1}(Color online) Temperature dependence of the resistivity of sample I in various magnetic fields. Open symbols are experimental data in magnetic fields from $0$ T to $9$ T at intervals of $0.5$ T. The red solid circles highlight the locations of the resistance minima. Dashed line represents the temperature dependence of the resistivity minima $\rho^* = [1+(m-1)^{-1}]\rho_0$ derived from Kohler's rule scaling MR $\sim (H/\rho_0)^m$ with $\rho_0$ being the experimental resistivity at zero field and $m = 1.92$ (see text for more discussion). The inset shows the Fermi liquid (FL) behavior $\rho_0 = A+BT^2$ for the resistivity at temperatures below $100$ K. Data were taken with magnetic fields applied along the c-axis of the crystal ($H\parallel c$) and current flowing along the a-axis.}
 \end{figure}

\begin{figure}
 \includegraphics[width=0.45\textwidth]{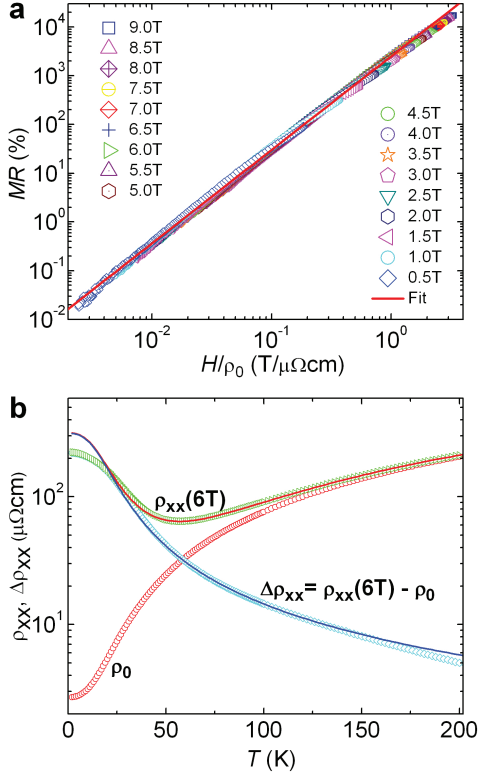}%
 \caption{\label{fig:fig2}(Color online) Kohler's rule analysis of the temperature dependence of the resistivity. \textbf{a}, Kohler's rule scaling of data in Fig.\ref{fig:fig1}. The symbols are experimental data and solid line
 represents a fit to $\mathrm{MR} = 25(H/\rho_0)^{1.92}$. \textbf{b}, temperature dependence of the resistivity at $0$ T and $6$ T and their differences. The solid lines are fits to Eq.1 with $\alpha = 25$ [$\mu \Omega$cm$/$T]$^{1.92}$ and $m = 1.92$. We present data up to $200$ K for clarity since the MR at $T > 200$ K is small
 (MR $\leq 5\%$, the value at $200$ K and $9$ T, see Fig.S2a).}
 \end{figure}

On the other hand, as presented in Fig.\ref{fig:fig2}a, all the data in Fig.\ref{fig:fig1} can be scaled onto a straight line when plotted as MR $\sim H/\rho_0$. That is, the temperature dependence of the magnetoresistance of this
sample follows the Kohler's rule:
\begin{equation}\label{eq:eq1}
\mathrm{MR} = \alpha (H/\rho_0)^m
\end{equation}
with $\alpha[=25(\mu\Omega\mathrm{cm}/\mathrm{T})^{1.92}]$ and $m[=1.92]$ being constants. This scaling behavior is valid for all the measured samples, as demonstrated in Fig.S3 for the data from sample II.

\begin{figure}
 \includegraphics[width=0.45\textwidth]{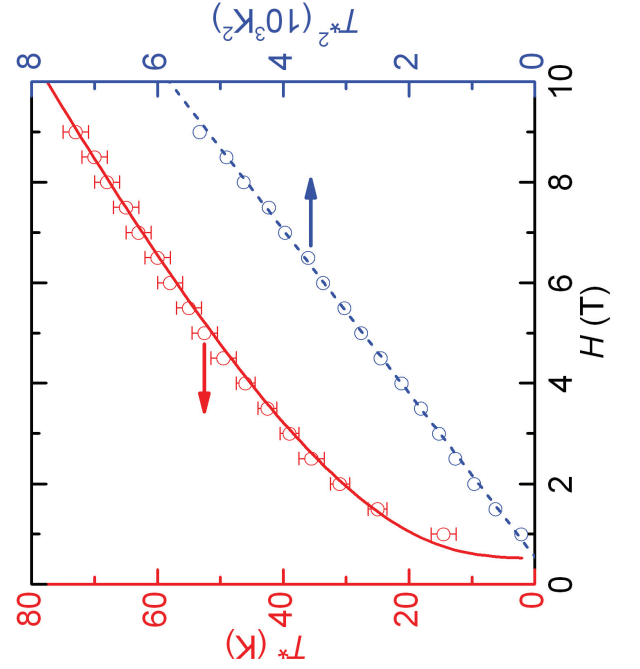}%
 \caption{\label{fig:fig3}(Color online) Magnetic field dependence of $T^*$ for sample I. The red solid line is a fit to $\rho_0(T^*)=H[\alpha (m-1)]^{1/m}$ and the dashed blue line represents ${T^*}^2 \sim H-0.5$.}
 \end{figure}

In order to showcase how Eq.\ref{eq:eq1} can lead to the remarked turn-on behavior shown in Fig.\ref{fig:fig1}, we replot the $0$ T and $6$ T resistivities as well as their difference $\Delta\rho_{xx}=\rho_{xx}(T,6)-\rho_{xx}(T,0)$ in Fig.\ref{fig:fig2}b. It clearly shows that the resistivity of a sample in a magnetic field consists of two components, i.e. $\rho_0$ and $\Delta\rho_{xx}$, with opposite temperature dependencies. In fact, Eq.\ref{eq:eq1} can be re-written as:
\begin{equation}\label{eq:eq2}
\rho_{xx}(T,H) = \rho_0+\alpha H^m/\rho_0^{m-1}
\end{equation}
The second term is the magnetic-field-induced resistivity $\Delta\rho_{xx}$, which is inversely proportional to $1/\rho_0$ (when $m = 2$) and competes with the first term when temperature is changed, resulting in a possible minimum at $T^*$ in the temperature dependence of the total resistivity. In fact, we can conveniently derive the $T^*(H)$ from Eq.\ref{eq:eq2} using $d\rho_{xx}(T,H)/d T = 0:\rho_0(T^*) = H[\alpha(m-1)]^{1/m}$. As
given in Fig.\ref{fig:fig3} for sample I, this relation (red solid line) correctly describes the experimental data (red open circles). The data and the fit in Fig.\ref{fig:fig3} also indicate the existence of a critical magnetic
field beyond which a resistivity minimum in the $\rho_{xx}(T,H)$ curve can occur. In fact, the $T^*(H)$ relation in our samples can also be described as $T^* \sim (H-H_c)^{1/2}$ (blue circles in Fig.\ref{fig:fig3}), consistent
with those observed in graphite and bismuth.\cite{R34} The physics behind it is simple: the resistivity minima occur in the Fermi liquid state\cite{R38} in which the temperature dependence of the zero-field resistivity follows $\rho_0 = A+BT^2$, as demonstrated in the inset of Fig.\ref{fig:fig1} for sample I. In that case, we have $T^* \sim (H-H_c)^{1/2}$ with $H_c = A[\alpha(m-1)]^{-1/m} \approx A/\alpha^{1/2}$ (for $m \approx 2$). For a quick estimate we replace $A$ with the resistivity value ($\approx 2.7$ $\mu\Omega$cm) obtained at lowest temperatures in zero field from Fig.\ref{fig:fig1}, resulting in $H_c \approx 0.52$ T, which is consistent with the experimental observation of the absence of a minimum at $0.5$ T in Fig.\ref{fig:fig1} and $H_c = 0.5$ T revealed by fitting the data (open blue
circles) in Fig.\ref{fig:fig3} using $T^* \sim (H-H_c)^{1/2}$ (dashed blue line). Remarkably, Eq.\ref{eq:eq2} also predicts an extremely simple temperature dependence for the resistivity $\rho^*$ at the minimum of the $\rho_{xx}(T,H)$
curve: $\rho^* = [1+(m-1)^{-1}]\rho_0$, i.e. $\rho^* \approx 2\rho_0$ since $m \approx 2$, as presented by the red dashed lines in Fig.\ref{fig:fig1}
for sample I and Fig.S3a for sample II. That is, the Kohler's rule Eq.\ref{eq:eq1} can quantitatively predict the temperature dependence of the total resistivity, including the resistivity minima and the astonishing up-turn at low temperatures in WTe$_2$, excluding the possible existence of a metalinsulator transition. The observed $T^* \sim (H-H_c)^{1/2}$ relationship is in fact an indication that the turn-on behavior occurs in the Fermi liquid state.

Originally, the Kohler's rule was developed to account for the magnetoresistance in metals, in which the magnitude of the magnetic field in theoretical derivation occurs always in the combination of $H\tau$ where $\tau$ is the relaxation time and related to $\rho_0$ through $\rho_0 = 1/n_ee\mu_e = m^*/n_ee^2\tau$ with $\mu_e$, $m^*$ and $n_e$ being the mobility, effective mass and density of conduction
electrons, respectively. Since $m^*$ is typically assumed to be temperature independent, Kohler's rule will be valid if the density $n_e$ is a constant. Kohler's rule plots in XMR materials showed both agreement\cite{R13,R14,R40,R44} and disparity.\cite{R17,R40,R44} As shown in Fig.S4, the $\mathrm{MR}(H)$ in our WTe$_2$ follows Kohler's rule Eq.\ref{eq:eq1} well, except for the data at very low temperatures. The derived form of the Kohler's rule (Eq.\ref{eq:eq2}) indicates that the temperature dependence of the measured resistivity in a
fixed magnetic field is solely determined by $\rho_0(T)$, because $\alpha$ and $m$ are temperature insensitive. Since $\rho_0(T)$ is inversely proportional to the temperature dependence of the mobilities $\mu_{e,h}(T)$,\cite{R45} Eq.\ref{eq:eq2} also reveals that the turn-on behavior in XMR materials originates from the strong
temperature dependence of the high mobilities of the charge carriers.

\begin{figure}
 \includegraphics[width=0.45\textwidth]{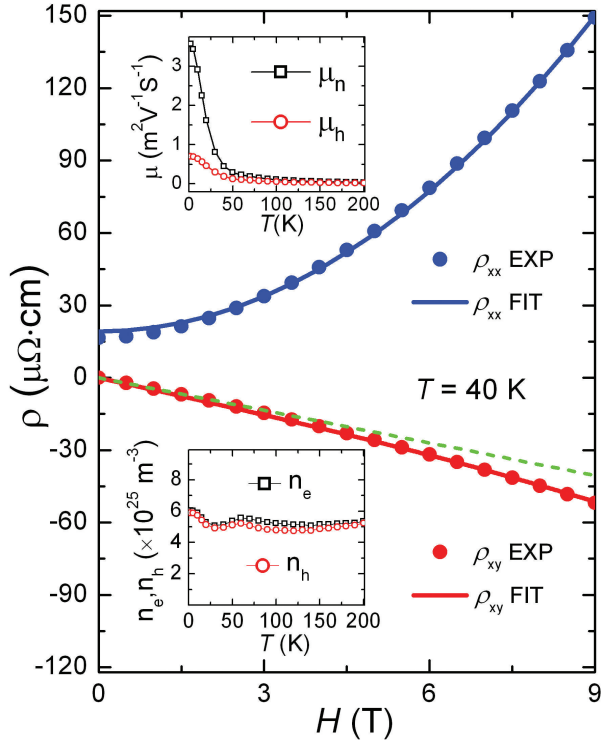}%
 \caption{\label{fig:fig4}(Color online) Direct fits to the two-band model represented by Eq.S1 for both the longitudinal ($\rho_{xx}$) and Hall ($\rho_{xy}$) resistivities of sample I. The green linear is a guide for the eyes. The upper and lower insets show the derived mobilities and densities of the electrons and holes.}
 \end{figure}

Although Kohler's rule is phenomenological, Eq.\ref{eq:eq1} with $m = 2$, i.e. $\mathrm{MR} \sim (H/\rho_0)^2$ can be derived from a two-band model for perfectly compensated systems.\cite{R45,R46} In WTe$_2$ the densities of electrons and holes are believed to be perfectly compensated.\cite{R15,R16} However, the exponent $m$ in Eq.\ref{eq:eq1} is not precisely $2$ and there exists detectable disparities between the experimental data at low temperatures and the fits to Eq.\ref{eq:eq1} in Fig.\ref{eq:eq2}. Besides experimental errors and possible electronic structure changes at low temperatures,\cite{R38,R39,R40,R41,R42} such a misfit, though not very significant, may indicate that the densities of the two types of charge carriers are not precisely equal. In fact, as shown in Fig.\ref{fig:fig4}, $\rho_{xy}(H)$ also deviates from a perfect linear behavior, revealing that $n_e \neq n_h$, i.e. the third term in the denominator of Eq.S1 cannot be completely neglected. Direct fits with Eq.S1 to both the $\rho_{xx}(H)$ and $\rho_{xy}(H)$ data presented in Fig.\ref{fig:fig4}, show that $n_e$ and $n_h$ do not change
significantly with temperature and have values of $(4.6-6)\times 10^{25}$ m$^{-3}$. The densities of the electrons and holes differ by $\sim 2-9\%$, depending on the temperature. The values are consistent with those derived from quantum oscillation measurements, which give $n_e = 6.64\times 10^{25}$ m$^{−3}$ and $n_h =6.9\times 10^{25}$ m$^{−3}$ at $0.59$ K.\cite{R47} These results also indicate that the electronic structure changes revealed by
$\mathrm{MR}$ anisotropy\cite{R39} and angle-resolved photoemission spectroscopy (ARPES) measurements\cite{R38} do not significantly modify either the temperature dependence or the absolute values of the carrier densities, although a slight abnormality is observed at $\sim 70$ K.

The above discussion indicates that Eq.\ref{eq:eq1} can also account for the temperature and magnetic field dependences of the resistivity for systems with imperfect compensation of the electron and hole densities. That is, Eq.\ref{eq:eq1} can fit the experimental data with negligible deviations if the first two terms in the denominator in Eq.S1 dominates. Since the third term in the denominator of Eq.S1 contains the product ($\mu_e \mu_h)$ of the two mobilities $\mu_e$ and $\mu_h$, its value should decrease faster than the first two terms with decreasing mobilities. Thus, Eq.\ref{eq:eq1} will be applicable for systems with a large difference in densities of electrons and holes, if either or both of the mobilities are small. In this case we can generalize the $\rho_{xx}(T,0)$ in Eq.S2 and $\alpha$ in Eq.\ref{eq:eq1} to be $\rho_{xx}(T,0) = [e\mu_e(n_e+\kappa n_h)]^{-1}$ and $\alpha =\kappa[e(n_e +\kappa n_h )]^{−2}$ , respectively.\cite{R45}

The applicability of Eq.\ref{eq:eq1} to non-compensated systems with low mobilities extends the importance of this work beyond understanding the magentoresistance in XMR materials: Eq.\ref{eq:eq1} also enables us to shed light on the magnetoresistances observed in non-XMR materials, which have mobilities\cite{R48} typically orders of magnitudes smaller than those in XMR materials. For example, the magnetoresistance in the normal state of cuprate superconductors was found to be only few percents at magnetic fields up to $30$ T and follows $\mathrm{MR} = \varepsilon H^2$, with $\varepsilon\propto T^{−4}$. \cite{R49} Without getting into details on possible bands of the charge carriers, Chan et al.\cite{R49} elucidate such temperature and magnetic field dependencies with the Kohler's rule, i.e. $\mathrm{MR} = f[H/\rho_0] \propto [H/\rho_0)]^2$
and $\rho_0 \propto T^2$ for the Fermi liquid normal state. Clearly, these $\mathrm{MR}$ behaviors are direct outcomes of Eq.\ref{eq:eq1}, revealing the two-band nature of the charge carriers in cuprate superconductors in which
the densities of holes and electrons can be controlled by doping.\cite{R49,R50}

Although Eq.\ref{eq:eq2} is aimed to clarify the so-called ‘metal-insulator transition’ in the temperature dependence of the magnetoresistance, it can easily account for the absence of the low temperature up-turn in $\rho_{xx}(T,H)$ of non-XMR materials in a fixed magnetic field: the high charge carrier densities ($n_e$ and/or $n_h$) lead to small $\alpha$. Together with the large residual resistance $A$, it can result in a large $H_c$ ($\approx A/\alpha^{1/2}$ in the Fermi liquid state) exceeding the magnetic fields available in a typical laboratory or a small $T^* [\sim (H-H_c)^{1/2}]$ that is beyond the experimentally
accessible temperature range.\cite{R44} In Fig.S5 we present the calculated $\rho_{xx}(T,H)$ curves using $\rho_{xx}(T,0)$ and $m$ of sample I while changing the value of $\alpha$ to demonstrate that the same Kohler's rule can lead to different temperature behavior: Fig.S5a shows that the turn-on temperature
behavior occurs only at $H > 2$ T when the $\alpha$ value decreases to $2.5$ ($\mu\Omega\mathrm{cm}/\mathrm{T})^{1.92}$; Fig.S5b indicates that in our experimentally accessible magnetic field of $9$ T, no turn-on behavior can be
observed if $\alpha = 0.15$ $(\mu\Omega\mathrm{cm}/\mathrm{T})^{1.92}$. Eq.\ref{eq:eq1} also implies that Kohler's rule will be violated if $\alpha$ is
temperature dependent. In this case one can obtain information on the temperature dependence of the charge carrier densities from Eq.\ref{eq:eq1} (using $m = 2$) with the measured $\rho_{xx}(T,H)$ through $\alpha=\rho_{xx}(T,0)\Delta\rho_{xx}(T,H)/H^2$.

In summary, we demonstrated that the Kohler's rule can account for the turn-on temperature behavior of the resistance in WTe$_2$, which seemingly looks like a metal-insulator transition. Based on the Kohler's rule scaling we could obtain the same magnetic field dependence of the turn-on temperature $T^* \sim (H-H_c)^{1/2}$, which was earlier considered as evidence for a metal-insulator
transition. We found a simple temperature dependence for the resistivity $\rho^* \approx 2\rho_0$ at the minimum of the $\rho_{xx}(T,H)$ curve. These results unambiguously demonstrate that the turn-on temperature behavior is not indicative of a metal-insulator transition but in fact of a high-quality and low charge carrier density sample (small residual resistivity, high mobilities, and large
residual resistance ratio) following Kohler's rule in a magnetic field. They also indicate that the electronic structure changes revealed by $\mathrm{MR}$ anisotropy and ARPES may not contribute to the
turn-on behavior. Our work not only resolves the long-time mystery of the turn-on temperature behavior in XMR materials but also provides a general route to understand the temperature behavior of measured resistances in both XMR and non-XMR materials.

\begin{acknowledgments} This work was supported by the U.S. Department of Energy, Office of Science, Basic Energy Sciences, Materials Sciences and Engineering Division. Use of the Center for Nanoscale Materials, an Office of Science user facility, was supported by the U. S. Department of Energy, Office of Science, Office of Basic Energy Sciences, under Contract No. DE-AC02-06CH11357. L.R.T. and Z.L.X. acknowledge NSF Grant No. DMR-1407175. The work at Tulane University was supported by the NSF under Grant DMR-1205469 and Louisiana
Board of Regents under grant LEQSF(2014-15)-ENH-TR-24.
\end{acknowledgments}

\bibliography{MITWTe2PrbV2}

\end{document}